\documentclass[namedreferences]{solarphysics}

\usepackage[hyperref,optionalrh,showbiblabels]{spr-sola-addons}
\usepackage{graphicx}
\usepackage{color}

\chardef\us=`\_

\begin{document}
\begin{article}
\begin{opening}

\title{Two New Methods for Counting and Tracking the Evolution of Polar Faculae}

\author[addressref={caltech,gsfc}, corref, email={berylha@caltech.edu}] {\inits{B.}\fnm{B.}~\lnm{Hovis-Afflerbach}\orcid{0000-0002-9967-2725}}
\author[addressref=gsfc, email={William.D.Pesnell@NASA.gov}] {\inits{W.D.}\fnm{W.~Dean}~\lnm{Pesnell}\orcid{0000-0002-8306-2500}}

\address[id=caltech]{California Institute of Technology, Pasadena CA 91125, USA}
\address[id=gsfc]{NASA Goddard Space Flight Center, Greenbelt MD 20771, USA}

\runningauthor{Hovis-Afflerbach and Pesnell}
\runningtitle{Evolution of Polar Faculae}

\begin{abstract}

Polar faculae are the footpoints of magnetic field lines near the Sun’s poles that are seen as bright regions along the edges of granules. 
The time variation in the number of polar faculae has been shown to correlate with the strength of the polar magnetic field and to be a predictor of the subsequent solar cycle.
Due to the small size and transient nature of these features, combined with different techniques and observational factors, previous counts of polar faculae differ in magnitude. 
Further, there were no scalable techniques to measure the statistical properties of the faculae, such as the variation of the facular lifetime with time or solar activity. 
Using data from the Helioseismic and Magnetic Imager (HMI) onboard the Solar Dynamics Observatory (SDO), we present two new methods for tracking  faculae and measuring their properties. 
In the first, we calculate the pixel-by-pixel standard deviation of the HMI continuum intensity images over one day, visualizing the faculae as streaks. The lifetime of the facula is found by dividing the angular length of the streaks by the latitude-dependent rotation rate.
We apply this method to the more visible pole each day for a week every six months, from September 2010 to March 2021.
Combining all of the measured facular lifetimes provides a statistical distribution with a mean of 6.0 hours, a FWHM of 5.4 hours, and a skew towards longer lifetimes, with some faculae lasting up to 1 day. 
In the second method, we overlay images of the progressive standard deviation with the HMI magnetogram to show the close relationship between the facular candidates and the magnetic field. 
The results of this method allow us to distinguish between motion due to the Sun’s rotation and “proper motion” due to faculae moving across the Sun’s surface, confirming that faculae participate in convective motions at the poles. 
Counts of polar faculae using both methods agree with previous counts in their variation with the solar cycle and the polar magnetic field. 
These methods can be extended to automate the identification and measurement of other properties of of polar faculae, which would allow for daily measurements of all faculae since SDO began operation in 2010.
\end{abstract}

\end{opening}

\section{Introduction}\label{sec:intro}

Faculae are bright spots in the solar photosphere where the solar magnetic field goes through the surface.
Several models have been proposed to explain the observed properties of faculae, which include broad-band emission at a temperature above the effective temperature of the Sun, the association with the magnetic field, and the variation in brightness with distance from the center of the disk. 
The ``hot wall'' model poses that faculae are depressions in the photosphere caused by the concentration of magnetic flux, and are bright because of the increased temperature of the wall of this depression \citep{Spruit_1976, Spruit_1977, keller_origin_2004}.
The ``hillock'' model poses that faculae are uplifted hillocks in the photosphere that have different contrast when seen from different viewing angles \citep{Schatten1986}.
The ``hot cloud'' model offers an explanation that faculae are hot clouds above the photosphere, held in place by the magnetic field \citep{Knoelker1988, Eker2003}. 

In this work, we focus on measuring the properties of polar faculae (PFe), which we assume means above a heliographic latitude of $70^\circ$.  Specifically, we investigate the number and lifetime of PFe that are unipolar footpoints of open magnetic fields. As they stay near the limb at all times, the evolution of PFe can be followed across the disk.

The polarity of the magnetic field in PFe tends to match that of the dominant polar magnetic field \citep{homann_spectro-polarimetry_1997}. The variation in their number has been shown to be correlated with the strength of the polar magnetic field \citep{1991ApJ...374..386S,Sheeley:2008ApJ...680.1553S, munoz-jaramillo_calibrating_2012} and has been described as a precursor prediction of the amplitude of the succeeding sunspot cycle by \citet{priyal_polar_2014}. 
The number of PFe at solar minimum was used as a precursor prediction of the amplitude of Solar Cycles 21 \citep{Schatten:1978GeoRL...5..411S}, 
22 \citep{makarov_polar_1989},
23 \citep{Schatten:1993lr}, 
24 \citep{Tlatov:2009SoPh..260..465T}, and 
25 \citep{janssens_prediction_2021}.

In general, PFe are coincident with strong upflows of 0.5--1~km s$^{-1}$ \citep{okunev_observations_2004, Blanco:2007AAp...474..251B}. The combination of a consistent upflow with an open magnetic field configuration means that PFe may be a source of the fast solar wind.

The lifetime of PFe is one feature that has been less well studied.  \cite{1955ZA.....38...37W} found that PFe lasted between a few minutes and six hours. Using similar techniques and data from October 1962, \citet{Cortesi:1978MitSZ.362.....C} found that the PFe lifetime was correlated with the size of the facula and derived lifetimes varying between 16 and 300~min. \citet{okunev_observations_2004} noted the lifetime of faculae exceeded their observation window of 1 hour, while \cite{Blanco:2007AAp...474..251B} state that PFe last from several hours to days. That there are few measurements of the facular lifetime is in part because the observations used for counting faculae are often too limited in time coverage (e.g., one observation per day) to observe PFe for their entire lifetimes. A study of the facular lifetime is increasingly relevant because of the work done by \cite{Attie:2015ApA...574A.106A} towards tracking small-scale features in the polar regions of the Sun to derive velocities. Features that last 4--6 hours are ideal for such tracking, and therefore a study of the lifetime of PFe could reveal whether PFe are ideal candidates.

The Helioseismic and Magnetic Imager \citep[HMI, ][]{2012SoPh..275..229S} onboard the Solar Dynamics Observatory \citep[SDO, ][]{2012SoPh..275....3P} provides frequent, high-resolution images and has sufficiently stable pointing to permit, for the first time, a simple statistical analysis to improve our observations of faculae as described in Section \ref{sec:methods}. Although faculae at any latitude can be analyzed with these methods, this work will concentrate on polar faculae.

Section \ref{sec:results} contains the distribution of facular lifetimes determined from these methods, as well as a comparison with the results of previous facular counts and the polar magnetic field. A discussion of these results is given in Section \ref{sec:discussion} and conclusions are given in Section \ref{sec:conclusions}.

\section{Data and Methods}\label{sec:methods}

We describe the SDO data we use, followed by our two methods. From these methods we measure several properties of PFe, including the position, presence of a magnetic field (reserving the strength for a future analysis), the lifetime, and the evolution of the number of PFe in time.

\subsection{SDO Data}

In this work we use two SDO HMI data series: continuum intensity and magnetogram images.

\begin{figure}
    \centering
    \includegraphics[width=\textwidth]{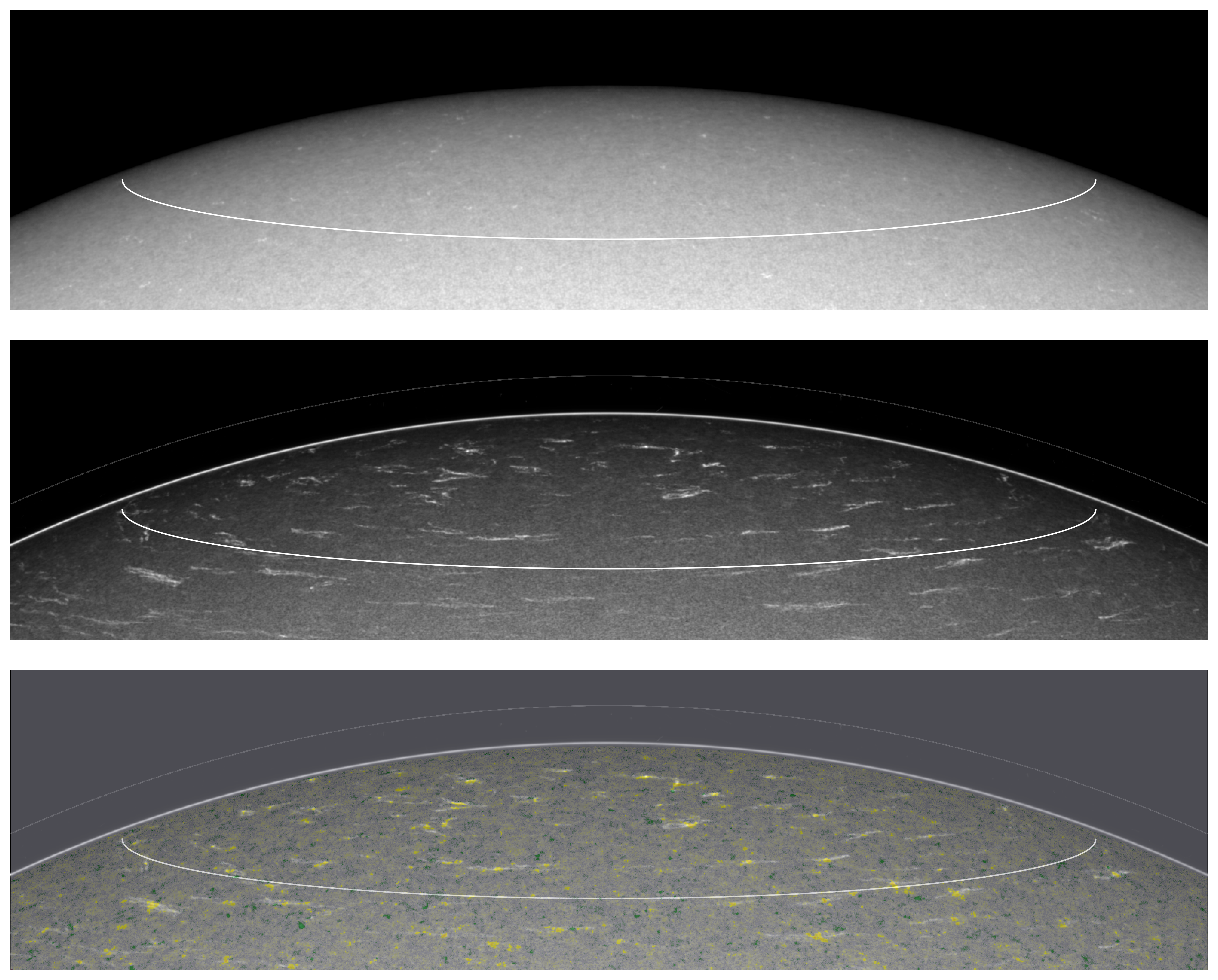}
    \caption{Examples of data and a daily standard-deviation (SD) image on September 2, 2010. (Top) An HMI${}_\mathrm{IC}$ image from the beginning of the day, scaled to increase contrast, (middle) a daily standard-deviation image, and (bottom) the daily standard-deviation image with 12:00Z HMI${}_\mathrm{M}$ overlaid. The white line denotes $70^\circ$ heliographic latitude in all of the panels. }
    \label{fig:example}
\end{figure}

HMI takes a full-disk continuum intensity image every 45~s, spectrally sampled in the wing of the Fe~\textsc{i} 6173.3~\AA\ absorption line  \citep{2016SoPh..291.1887C}, that is usable as a proxy for a white light image of the photosphere.
At this wavelength, polar faculae are visible as bright points in the photosphere.
We use the HMI.Ic\_720s series (HMI${}_\mathrm{IC}$) that is an average of the full-cadence data over 720~s (12~min), resulting in 120 images per day. This averaging process reduces the fluctuations caused by granules, which have a lifetime of 8--10 min, and the 5-min variations of the $p$-modes.
A portion of an example image from the HMI${}_\mathrm{IC}$ series is shown in the first panel of Figure~\ref{fig:example}.

The line of sight (LOS) magnetograms of the HMI.M\_720s series (HMI${}_\mathrm{M}$), which is also averaged over 12~min and is spatially registered with the intensity images \citep{2016SoPh..291.1887C}, are used for magnetic field data. Together, these two data series show faculae in the photosphere in visible light and provide information about their magnetic field.

\subsection{The First Method: Daily Standard-Deviation Image}

\begin{figure}
    \centering
    \includegraphics[width=.7\textwidth]{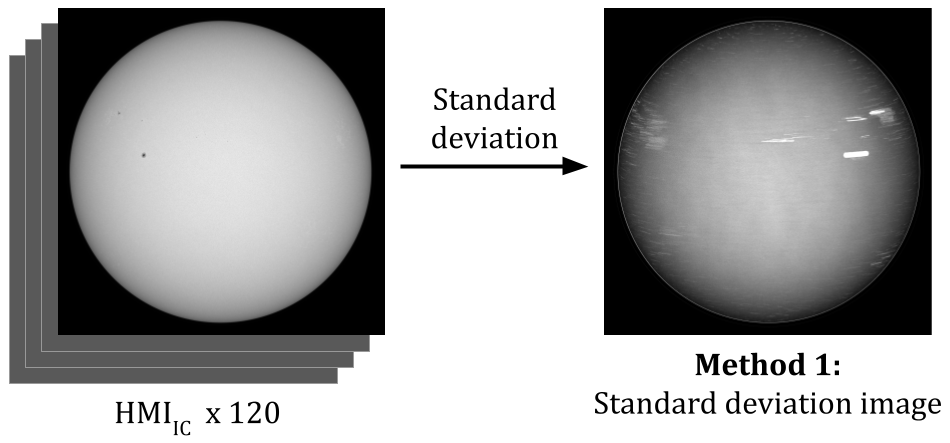}
    \caption{Schematic showing the process of the first method, which produces a standard-deviation image, for September 2, 2010. We take the pixel-by-pixel standard deviation of 120 images from the HMI${}_\mathrm{IC}$ series, which is sampled every 12 minutes over one day. In the standard-deviation image, faculae are seen as streaks due to the Sun's rotation, making them easy to identify and measure.}
    \label{fig:schematic1}
\end{figure}

As illustrated in Figure~\ref{fig:schematic1}, we calculate the standard deviation of each pixel in all 120 images in the HMI${}_\mathrm{IC}$ series over one day.
This method was initially developed to search for comet tails in EUV images \citep{2016ApJ...822...77B}, as it is sensitive to small, sequential changes in brightness even in the presence of significant noise. 
This method has some similarities with persistence mapping, a method that can be used for higher-contrast variations \citep{Thompson:2016ApJ...825...27T}.

A portion of a daily standard-deviation (SD) image produced using this method is shown in the second panel of Figure~\ref{fig:example}.
Due to the Sun’s rotation, the faculae appear as streaks moving across the Sun.
The brightest features in Figure~\ref{fig:schematic1} are active regions and their attendant faculae, which appear far from the poles at active latitudes and whose sunspot component can be visible far from the limb.
Though active regions appear dark in the original HMI${}_\mathrm{IC}$ images, the SD image shows the absolute value of the contrast, so both bright and dark features appear brighter in an SD image.

As SDO is a space telescope and has no seeing, we can confirm that the features in this standard-deviation image correspond to features on the Sun.
For a ground-based telescope, the standard-deviation image would always be dominated by the seeing, and this method would be impossible.

\subsection{The Second Method: Progressive Standard-Deviation Movie}

\begin{figure}
    \centering
    \includegraphics[width=\textwidth]{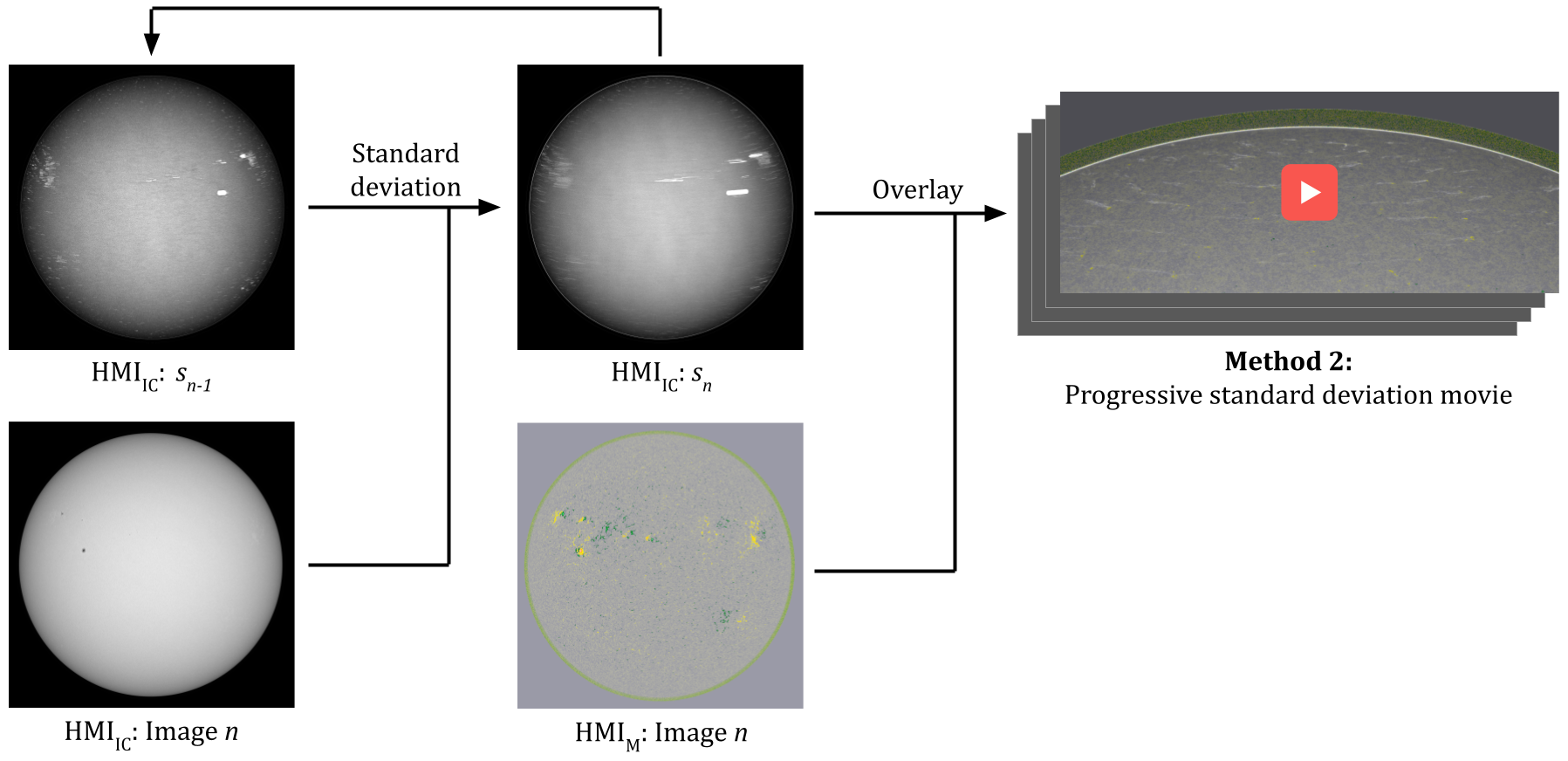}
    \caption{Schematic showing the process of the second method, which produces a progressive standard-deviation movie, for September 2, 2010. Again using the 12-min HMI${}_\mathrm{IC}$ series, at each time step we take the standard deviation up to that point recursively, then overlay the HMI magnetogram (HMI${}_\mathrm{M}$) for that time step. The resulting movie shows the evolution of the streaks visible in the first method, as well as their associated magnetic-field elements, and is included in \href{https://static-content.springer.com/esm/art\%3A10.1007\%2Fs11207-022-01977-8/MediaObjects/11207_2022_1977_MOESM1_ESM.mp4}{Supplemental Material}.}
    \label{fig:schematic2}
\end{figure}

Our second method, illustrated in Figure~\ref{fig:schematic2}, uses a recursive definition of the sample average ($\bar{x}_n$) and variance ($s_n^2$) given by Equations~\ref{eq:avg} and \ref{eq:var} \citep{Welford_1962}. The $N$ measurements are denoted $x_n$, for $n=1, \ldots, N$, and the recursion is started with the initial values $\bar{x}_1 = x_1$ and $s_1^2 = 0$. Values for $n=2, \ldots, N$ are then given by
\begin{eqnarray}
    \bar{x}_{n} & = & \frac{1}{n} \left[ (n-1) \bar{x}_{n-1} + x_{n} \right] \label{eq:avg} \\
    s_n^2 & = & \frac{n-2}{n-1} s_{n-1}^2 + \frac{n (x_n - \bar{x}_n)^2}{(n-1)^2}. \label{eq:var}
\end{eqnarray}

For each $n=2, \ldots, N$, we do the following:
\begin{enumerate}
    \item Using image $n$ of the HMI${}_\mathrm{IC}$ series data for the day, calculate the average image $\bar{x}_n$ according to Equation~\ref{eq:avg};
    \item Calculate the variance image $s_n^2$ according to Equation~\ref{eq:var};
    \item Take the square root of the variance image $s_n^2$ to produce the standard-deviation image $s_n$; and
    \item Overlay image $n$ of the HMI${}_\mathrm{M}$ series onto the standard-deviation image $s_n$.
\end{enumerate}

By concatenating the images produced from step 4 of this process for $n=2, \ldots, N$, we produce a movie that shows the evolution of the streaks visible in the first method.
Each streak -- each facula -- is associated with a magnetic-field element.
A movie for September 2, 2010 can be seen in \href{https://static-content.springer.com/esm/art\%3A10.1007\%2Fs11207-022-01977-8/MediaObjects/11207_2022_1977_MOESM1_ESM.mp4}{Supplemental Material}.

\subsection{Measuring Faculae Using Standard-Deviation Images}\label{subsec:measuring}

A daily standard-deviation image is used to count the number of faculae visible over an entire day.
In general, faculae appear as streaks in the daily SD images.
When counting streaks, we selected the endpoints of each streak and recorded their pixel positions.
As this was a preliminary exploration of this method intended for comparison with historical counts, which were primarily done by eye, we selected faculae by eye, without the use of any kind of automated selection method based on a threshold contrast.
The potential of an automated method is discussed at the end of Section \ref{subsec:further} and is a primary focus for further investigation.

The streaks are not always simple lines.
The magnetic field from the same day is therefore used to assist in identifying faculae. 
The 12:00Z magnetogram image from the HMI${}_\mathrm{M}$ series is first overlaid on the daily SD image.
An example of a combined image from September 2, 2010 is shown in the bottom panel of Figure~\ref{fig:example}.
The magnetogram overlay provides information about the magnetic-field element associated with each bright streak, which is used to determine to whether it is a PF or another phenomenon.

\begin{figure}
    \centering
    \includegraphics[width=\textwidth]{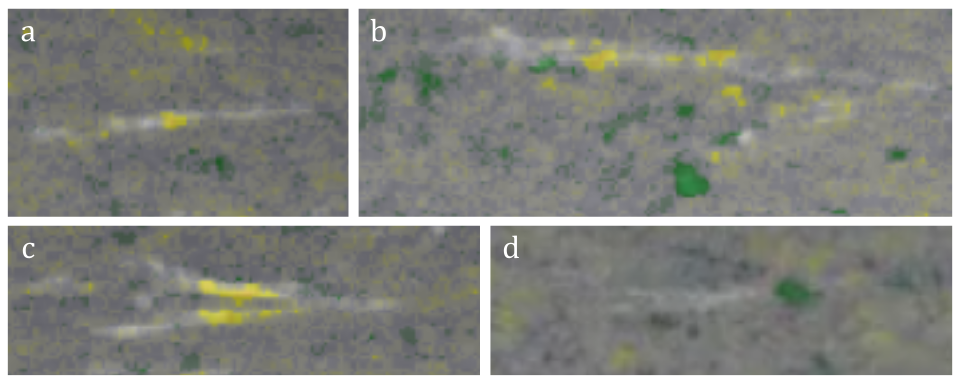}
    \caption{Typical examples from September 2, 2010 of features seen using the two methods: (a) a unipolar facula seen as a streak in the standard-deviation image and with the same polarity (negative) as the pole, (b) a multipolar region seen as a cluster of streaks and with both positive and negative polarity, (c) two streaks with latitudinal proper motion, and (d) a unipolar facula with opposite polarity to the pole. All of these examples have the same scale and can be seen in more detail in the progressive standard-deviation movie of September 2, 2010, included in \href{https://static-content.springer.com/esm/art\%3A10.1007\%2Fs11207-022-01977-8/MediaObjects/11207_2022_1977_MOESM1_ESM.mp4}{Supplemental Material}.}
    \label{fig:features}
\end{figure}

With this added information, we can distinguish between the more typical unipolar faculae we are studying and other features with different characteristics, such as multipolar regions, faculae with proper motion, and faculae with polarity that differs from that of the pole.
Examples of these different types of features as seen on September 2, 2010 are shown in Figure~\ref{fig:features}, and the latter two are discussed in greater detail in Section \ref{subsec:features}.

Most faculae look similar to the typical facula shown in Figure~\ref{fig:features}a.
In the progressive standard-deviation movie, each facula would be clearly associated with a magnetic-field element, and once this magnetic-field element fades, so does the facula, and the streak ends.

Some bright points in the poles do not fall within this classic definition of faculae but are instead seen as multipolar regions of magnetic field, such as the one shown in Figure~\ref{fig:features}b.
These points are not footpoints of an open magnetic field, but rather small, multipolar regions of a closed field.
Some previous studies of faculae, such as that done by \cite{makarov_polar_1989}, have included these regions in counts of polar faculae.
However, because we expect these multipolar regions to evolve and behave differently from unipolar faculae and because our method enables their distinction, we include only those faculae that are associated with unipolar magnetic-field elements in our analyses described below.

Some faculae have a dotted appearance.
Because facular brightness can vary with time, we assumed that these streaks were a single facula when counting.
In particularly ambiguous cases, the progressive standard-deviation movie can provide confirmation of whether a streak is a single facula or not.

We define and count polar faculae as those above $70^\circ$ latitude. \cite{1923MNRAS..84...96.} found a slight concentration of faculae at $70^\circ$ and determined it to be the boundary at which polar faculae differ from lower-latitude faculae in their appearance, lifetime, and relationship with sunspots and the solar cycle. The choice of $70^\circ$ was also used by, e.g.,  \cite{munoz-jaramillo_calibrating_2012}.

Due the inclination of the Sun's axis of rotation relative to the ecliptic, the solar north pole is slightly more visible from Earth in the fall, while the solar south pole is more visible in the spring. 
So that we always measure faculae from a constant angle of inclination from the Sun's equator, we measure faculae on each day from September 1--7 and from March 1--7, when the north and south poles, respectively, are most visible from Earth. We record faculae only on the more visible pole.

To reduce the subjectivity in the selection of data, the day for which faculae were to be counted was drawn at random from the list of days not already measured. While counting faculae on the image for that day, we were not aware of the date nor the position in the solar cycle.

The sign of the facular count is chosen to agree with the polarity of the polar magnetic field of the same hemisphere at that time.
For this work, we assume the reversal of polarity to occur in March 2014 for the south pole and in November 2012 for the north pole, as found by \cite{Sun:2015ApJ...798..114S}.

\subsection{Calculation of the Lifetimes of Polar Faculae}\label{subsec:calc_life}

Our methods allow us not only to count the number of PFe but to also  measure their lifetime.
We assume that the streaks are formed by faculae moving along a constant heliographic latitude ($\vartheta$) at a latitude-dependent rotational angular velocity ($\Omega[\vartheta]$).

We use the $\Omega[\vartheta]$ prescription from \cite{howard_spectroscopic_1970}, which is given by their Equation~1, using the values of $a$, $b$, and $c$ listed as ``All Data'' and ``Full Disk'' in their Table~I.
This equation with these values is reproduced in Equations~\ref{eq:hh1} and \ref{eq:hh2} below.
\begin{eqnarray}
    \Omega(\vartheta) & = & 2.78 \times 10^{-6} - 3.51 \times 10^{-7} \sin^2 \vartheta - 4.43 \times 10^{-7} \sin^4 \vartheta \mbox{ rad/sec} \label{eq:hh1} \\
& = & 13.8 - 1.74 \, \sin^2 \vartheta - 2.19 \, \sin^4 \vartheta \,\,\, {}^\circ/\mbox{day} \label{eq:hh2}
\end{eqnarray}

We use the same image used to count the faculae, which is described in Section \ref{subsec:measuring}, and record the pixel position of the beginning and the end of each streak.
We then convert these pixel coordinates to heliographic latitude and longitude using the Python packages \texttt{astropy.coordinates.SkyCoord} and \texttt{sunpy.coordinates.frames}.
By calculating the distance between these heliographic endpoints of each streak ($\Delta \phi$) and dividing by the rotational angular velocity, we calculate the lifetime of each facula as
\begin{equation}
    \tau = \Delta \phi/\Omega(\vartheta) \mbox{  sec}.
\label{eq:tau}
\end{equation}

Some faculae are very close to the limb and rotate into or out of view of SDO during the day.
It would be inaccurate to measure the facula's lifetime using only the part of the streak that is in view.
Therefore, when measuring the lifetime of PFe, we exclude all of those streaks that have one endpoint $\phi$ for which $|\phi| > 90^\circ$.
However, these streaks are still included in the facular count described in Section \ref{subsec:count}.

\section{Results}\label{sec:results}

\subsection{Distribution of Facular Lifetimes}

How long a facula lasts is a diagnostic of how it is formed or destroyed. The facular lifetime ($\tau$) can be investigated by measuring the lengths of the streaks in the standard-deviation images, as described in Section \ref{subsec:measuring}.

\begin{figure}
    \centering
    \includegraphics[width=\textwidth]{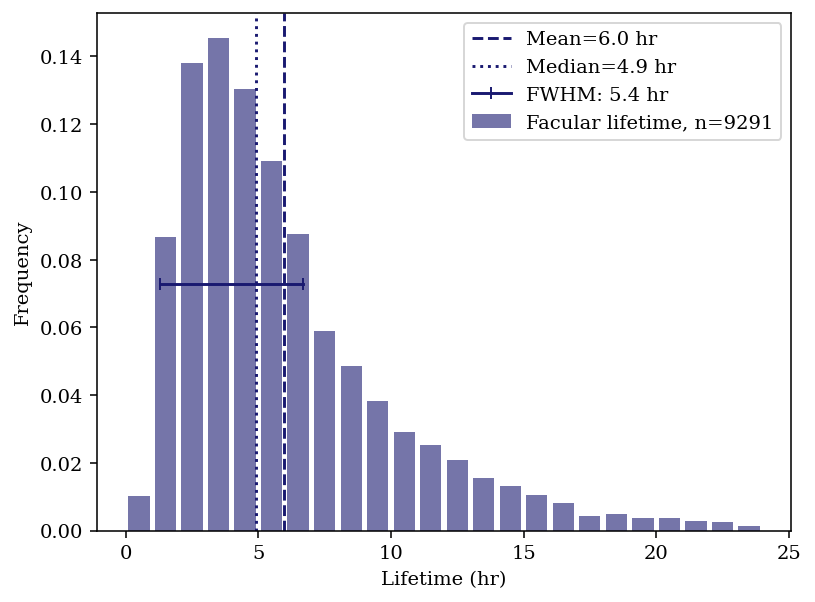}
    \caption{The distribution of lifetimes calculated for all measured faculae. The mean lifetime is 6 hours, while the median is 4.9 hours, and the full width at half-maximum is 5.4 hours, from 1.3 to 6.7 hours. The distribution is skewed to the right.}
    \label{fig:lifetime}
\end{figure}

The distribution of the measured lifetimes of 9291 PFe is shown in Figure~\ref{fig:lifetime}. The distribution has a mean of 6.0 hours, a median of 4.9 hours, and a full width at half-maximum of 5.4 hours. It skews to the right, as some faculae last up to a day.

Though this distribution resembles a Poisson distribution, an attempted fit of the data to a Poisson distribution did not produce useful results.
A Poisson distribution would predict a less significant tail on the right and therefore fewer faculae with longer lifetimes.
Some possible reasons for this discrepancy are discussed in Section \ref{subsec:further}.

As the data series used to make the standard-deviation image and movie, HMI${}_\mathrm{IC}$, are averaged over 12 min, any faculae that exist for less than 12 minutes will not be visible using these methods.
However, because the distribution of facular lifetime (Figure~\ref{fig:lifetime}) peaks at 3-4 hours, with fewer from 2-3 hours, fewer still from 1-2 hours, and almost none less than 1 hour, we consider it unlikely that there are many faculae with lifetimes this short.

This is in contradiction to \cite{1955ZA.....38...37W} who found two components of the population of PFe, one that had longer lifetimes and one with shorter lifetimes.
The population with longer lifetimes lasted around 240 or 160 minutes (using data from 1953 and 1954, respectively).
By contrast, the shorter-lived population had lifetimes around 6.3 to 18 minutes (again using data from 1953 and 1954).
These shorter-lived PFe would be poorly resolved in this study due to the 12-min averaging of the data.
This difference can be reconciled using the HMI.Ic\_45s series, which contains images taken every 45 seconds and is not averaged over a long time interval.
This data series should make it possible to find shorter-lived PFe using our methods.

In general, multipolar regions as described in Section \ref{subsec:measuring} tend to have longer lifetimes than the unipolar faculae reported here.
The exact measurement of the lifetimes of multipolar faculae is outside the scope of this work but should be possible using our methods.

\subsection{Facular Count over Solar Cycle 24}\label{subsec:count}

\begin{figure}
    \centering
    \includegraphics[width=\textwidth]{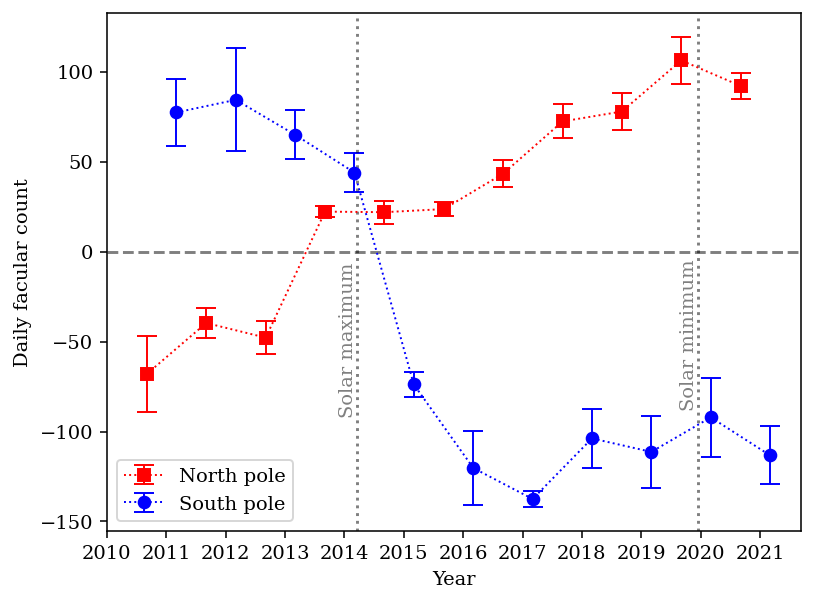}
    \caption{Averaged daily facular counts for September 1--7 and March 1--7 of each year, corresponding to the north and south poles, respectively. The measurements span September 2010, near the beginning of SDO operations, through March 2021. The sign of the measurement is chosen to agree with the polarity of the pole at that time. Solar maximum, in early 2014, and solar minimum, in late 2019, are indicated with vertical dashed lines.}
    \label{fig:counts}
\end{figure}

As described in Section \ref{subsec:measuring}, we count the number of faculae each day for one week every 6 months from September 2010, at the beginning of SDO’s operation, through March 2021.
Of the 154 days in this time frame, we measured faculae on 134 (87\%) of the days. The total number of faculae counted is 9375, an average of 70 faculae per day. A daily-averaged facular count is produced by averaging values from the seven days.
Variations in the daily-averaged facular count over Solar Cycle 24 are shown in Figure~\ref{fig:counts}.

The SDO launch in 2010 occurred too late to observe the solar minimum in late 2008, but we do observe the behavior of PFe during the next two extrema of the sunspot cycle. The number of polar faculae is small around solar maximum in 2014, when the polar magnetic field reversal takes place, and peaks around solar minimum in late 2019. 
This pattern is similar to that found in previous studies, such as \cite{Sheeley:2008ApJ...680.1553S} and \cite{munoz-jaramillo_calibrating_2012}.
However, our numbers tend to be higher than most previous daily facular counts.

These higher facular counts are in part a consequence of using HMI data.
As a space telescope, SDO and its observations are not affected by atmospheric seeing.
By contrast, ground-based observations of faculae such as those done by \cite{Sheeley:2008ApJ...680.1553S}, who used plates from Mount Wilson Observatory, and \cite{janssens_prediction_2021}, who counted faculae directly by eye, are affected by atmospheric seeing conditions, which can make it difficult to resolve fainter, low-contrast, faculae.

HMI takes full Sun images in high resolution.
In a previous study, \cite{munoz-jaramillo_calibrating_2012} counted PFe using continuum images from the Michelson Doppler Imager (MDI) onboard the Solar and Heliospheric Observatory (SoHO) for most of Solar Cycles 22 and 23.
Like HMI, MDI benefits from viewing the Sun in space, in this case from the Sun-Earth L1 Lagrange point.
However, MDI has a two-pixel resolution of $4.8''$, a factor of 4 lower than HMI's two-pixel resolution of $1.2''$.
Smaller faculae that are not resolved by MDI might be resolved by HMI, so any observation of faculae using HMI is likely to result in proportionally more faculae than seen by MDI.

The higher facular counts reported here are also a result of using the daily SD images.
While previous methods of counting faculae have relied on counting how many faculae are visible at a particular moment, the daily SD image allows the counting of faculae over an entire day by making all of those faculae visible in just one image.
Once more days have been counted, then a direct comparison of facular counts can be made with data from \cite{janssens_prediction_2021}, whose observations span a similar time range. All other long-term datasets have poor overlap with the SDO data.

\subsection{Comparison of Facular Count with WSO Polar Magnetic Field}

\begin{figure}
    \centering
    \includegraphics[width=\textwidth]{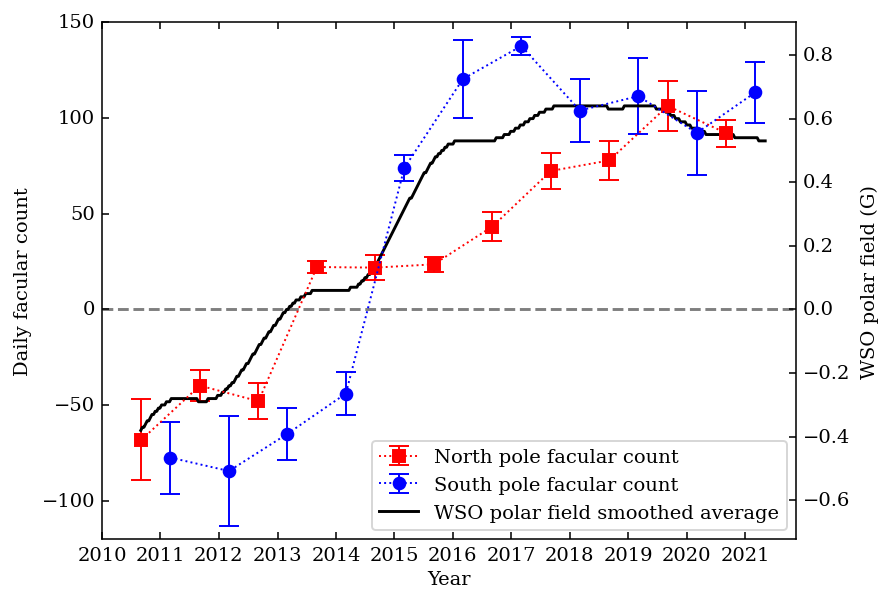}
    \caption{Averaged daily facular counts from Figure~\ref{fig:counts} compared with the smoothed average WSO polar magnetic field. The sign of the south pole facular count has been reversed compared with Figure~\ref{fig:counts} to better match $B_{pol}$.}
    \label{fig:wso}
\end{figure}

The number of PFe has been shown to track with the magnitude of the solar polar field \citep{1991ApJ...374..386S}. We demonstrate this relationship using the Wilcox Solar Observatory (WSO) measurements of the solar polar magnetic field ($B_{pol}$) recorded between September 2010 and March 2021. For our purposes the magnetic fields from the northern and southern hemispheres were averaged together to reduce the effect of the changing viewpoint caused by the Earth's orbital motion and the angle between the ecliptic and the solar rotation axis. This field is shown as the black line in Figure~\ref{fig:wso}.

To compare our facular counts with the WSO polar field, we overplot the counts from Figure~\ref{fig:counts}, reversing the sign of the south pole counts so as to better agree with the WSO polar magnetic field average.
The facular counts indeed follow the same pattern as the WSO polar field, with the black line tracking the polar field smoothed average falling along or between the red and blue lines tracking the north and south pole facular counts, respectively.
Though there is a good visual correlation between our facular counts and $B_{pol}$, our facular counts are not conducted at a sufficiently frequent interval to perform a quantitative comparison between the two data sets.
This correlation agrees with previous results from \cite{1991ApJ...374..386S} and \cite{munoz-jaramillo_calibrating_2012}.

The PF counts in the northern hemisphere show the effects of the stagnation of the reversal of the northern polar field. Unlike the southern hemisphere, which moved smoothly through 0~gauss at solar maximum, the northern hemisphere stayed near 0~gauss for over a year as a surge of old-cycle magnetic field entered the northern polar region \citep{2016STP.....2a...3M}. Figure~\ref{fig:counts} also shows that the PF counts in the northern hemisphere not only remained constant from September 2014 through September 2016 but had a smaller standard deviation compared to other parts of Solar Cycle 24.

\section{Discussion}\label{sec:discussion}

\subsection{Benefits of the Two New Methods}

Both the standard-deviation image method and the progressive standard-devia-\\tion movie method provide new ways to visualize PFe and better understand their properties and evolution.
They provide a facular count that follows the same pattern as previous methods.
Further, they provide an improvement over previous methods in a number of ways.

When studied using these methods, faculae become clearer and more noticeable, as the standard-deviation method increases the contrast and a streak is easier to notice and study than a single point. Faculae with weaker contrast may be more visible as a streak due to the correlated brightening along a streak.
Because the method tracks the facula for many images, it can confirm that this feature, which was previously only seen as a single point, is in fact a facula and not merely statistical brightening.

These methods make it easier to view and study the entire evolution of a single facula because each streak maps the facula over its entire lifetime.
While the first method, the daily standard-deviation image, enables visualization of the entire life of a facula at once, it does so at the expense of most time evolution, which is removed to flatten all the day's images into one.
The second method, the progressive standard-deviation movie, improves over the first method in that it shows the evolution of this streak in time and in that it overlays the corresponding LOS magnetic-field strength.
Though the second method makes it easier to associate faculae with magnetic-field data, it is easier to measure the positions of faculae using the first.

Previously, to count the number of faculae on one day, it was necessary to look at many images taken over the course of the day and record the number on each frame separately.
However, because the daily SD image contains all of the faculae for a single day, it is just as easy to count an entire day's worth of faculae as it has previously been to count the faculae visible at just one moment.

This also increases the ease with which one can observe and count multiple days' worth of faculae.
The daily counts, as seen in Figure~\ref{fig:counts}, have standard deviations ranging from approximately 3 to 30 faculae.
This indicates that a daily count of PFe can have substantial fluctuations, and it is not uncommon for the number of faculae to differ by up to 30 from day to day.
Multiple days' worth of counts, therefore, are more representative of the behavior of faculae at a given time in the year than only a single day's worth.

These methods also address the problem of inadvertently double counting faculae.
It was previously necessary to manually trace each facula from frame to frame as it moves across the Sun to ensure that no facula was double counted.
Our methods prevent double counting because each facula appears as a single streak.

Further, to measure the lifetime of a facula, one would have to note its first appearance, trace it across many time frames, and note its  disappearance. This is shown explicitly in Fig.~3 of \citet{1955ZA.....38...37W} and Fig.~1 of \citet{Cortesi:1978MitSZ.362.....C}.
Our methods expedite this process, as the start and end points of each streak are clear and easy to measure.  
Calculating the lifetime of the facula then requires analyzing only one image, rather than many.

Overlaying the magnetogram, as is done when counting faculae using the first method and at each timestep when producing the progressive standard-deviation movie in the second method, allows us to distinguish between unipolar faculae and multipolar regions.
As the behavior of multipolar regions differs from that of unipolar faculae, we study the two phenomena separately.
This magnetogram overlay, especially over time as done in the second method, makes this possible.

\subsection{Other Observed Features}\label{subsec:features}

Our analysis reveals other properties of PFe, examples of which are shown in the bottom half of Figure~\ref{fig:features}. We have assumed streaks to follow lines of constant latitude, as they would if the motion of faculae as observed were due solely to the Sun's rotation.
However, some faculae exhibit ``proper motion'' due to the motion of the faculae on the Sun, not just the rotation of the Sun itself.
We see motions in both directions, indicating this motion is solar in origin and not caused by the data-reduction technique.
In some cases, such as Figure~\ref{fig:features}c, two nearby faculae with latitudinal proper motion can form a ``V'' shape, making this proper motion even clearer.

For our measurements of facular lifetime, we have assumed a constant latitude $\vartheta$ when calculating $\Omega[\vartheta]$ and a change in longitude $\Delta \phi$ that is due only to this rotation rate.
For cases in which a facula has some proper motion in latitude, we have used its mean latitude for the calculation of the solar rotation rate.
Our methods make visible proper motion in latitude, as many streaks can be seen not to follow lines of constant latitude, but it is not as clear from either method whether a facula has longitudinal proper motion.
Such motion would have some effect on the calculation of a facula's lifetime.
However, because the latitudinal proper motions we see occur in both directions (northward and southward), we find it likely that any longitudinal proper motions would do the same (eastward and westward).
Any uncertainty introduced by these assumptions, then, would not introduce systematic bias or affect the distribution of facular lifetimes in Figure~\ref{fig:lifetime}.

As found by \cite{homann_spectro-polarimetry_1997}, the magnetic field in most faculae has the same polarity as the polar magnetic field.
However, a few have the opposite polarity, such as the one shown in Figure~\ref{fig:features}d.
In this case, the north pole has negative (yellow) polarity, but the facula shown has positive (green) polarity.
These features are far less common than faculae with the same polarity as the pole, and we rarely find more than one or two per day.

The brightness of a streak does not remain constant during the evolution of a facula. It is possible that a facular region represents a part of the solar atmosphere that brightens and dims on one timescale and evolves as a facula on a longer timescale. This would agree with the conclusions of \cite{Hirayama:1978PASJ...30..337H} using faculae in the active latitudes. PFe provide a way to study this possibility as the latitudes that they populate are high enough that faculae can be seen for almost an entire disk passage.
If faculae appear and disappear in the same spot that rotates along with the Sun, it could confirm whether faculae last longer than a single brightening.
This could be investigated using data from solar maximum, when there are fewer PFe and it is easier to track a single facula.

\subsection{Other Observations and Further Investigation}\label{subsec:further}

This report is only a first look through the data using this technique. Further analysis may provide insight by accounting for a few effects that we noticed during this study.

As seen in Figures \ref{fig:example}, \ref{fig:schematic1}, and \ref{fig:schematic2}, the HMI${}_\mathrm{IC}$ images are darker near the limb, which then causes the SD image to also be darker near the limb.
This effect can be removed by subtracting the limb-darkening profile from the HMI${}_\mathrm{IC}$ images, then creating a SD image from these images without limb darkening.
A comparison of counts with and without limb darkening resulted in only a small change in the number of faculae counted.
While SD images made from solar images with the limb-darkening profile subtracted are optimal for future use with this method, SD images that still contain limb darkening can be useful for comparison with historical counts, as many of those were done using observations that did exhibit limb darkening.

In the progressive standard-deviation movie, some streaks are bright when they first appear, then fade over the course of the day as more images are added to the standard deviation. However, we do not find any instances of streaks that fade so much that they are no longer discernible in the standard-deviation image, so the facular count is not affected. Additionally, though a streak may fade somewhat, its length and morphology remain the same. This means that the measured lifetime is also not affected. To reduce fading, the standard-deviation image could be made using every third or fourth HMI${}_\mathrm{IC}$ image, for a total of only 40 or 30 images over the entire day, respectively, with a reduced accuracy in the derived lifetime.

There are cases in which streaks overlap, making it difficult to discern whether a streak is two separate faculae or a single, longer facula.
A bias towards miscounting two streaks as one longer one would result in  under-counting the actual number of faculae and over-estimating the lifetime of a few faculae, making the right-hand tail of the distribution in Figure~\ref{fig:lifetime} more prominent.
Because of this source of uncertainty in the lifetime measurement, the  number of faculae with longer lifetimes may be systematically too high and would likely shift towards shorter values with re-measurement.
It is possible that with further analysis and correction of this source of uncertainty, the histogram of lifetimes may better resemble a Poisson distribution. Because there are few long-lived PFe, this would not be a large change.

It is inevitable that some faculae will start on one day and end on the next.
As we only measure the faculae as visible on the day of measurement, between 00:00Z and 23:59Z, any faculae that either begin on the previous day and end on the day of measurement or begin on the day of measurement and end on the following day will be cut off and their streaks will appear shorter than their true length.
This would tend to bias the distribution of facular lifetime towards the shorter end.

To estimate the impact of this effect on the distribution of facular lifetimes, we compared the distribution of facular lifetime produced by combining measurements of two standard-deviation images from two subsequent days with the distribution produced from measurements of a single standard-deviation image run over the two days.
We found that the distributions lined up fairly well, and any uncertainty introduced by the cut-off at one day is small.
Further, increasing the length of time covered by a standard-deviation image also increases the number of faculae which overlap, making the measurement of the position and lifetime more difficult and increasing the uncertainty of the measurements as described above.
Because the uncertainty introduced by the cut-off is small and because almost all faculae last less than one day, we find one day to be an appropriate length of measurement.

After creating a daily SD image, we overlay the magnetogram from 12:00Z on the date of measurement, as described in Section \ref{subsec:count}.
Some faculae exist only before or after this time, so while they are visible as streaks in the daily SD image, their associated magnetic-field element may not be visible.
Because we use the overlain magnetogram to distinguish between unipolar faculae, which we count, and multipolar regions, which we do not count, any streaks that exist only prior to or after 12:00Z and therefore have ambiguous polarity become difficult to distinguish.
In most cases, we assume that these streaks with ambiguous polarity are unipolar and count them as faculae.
An exception occurs when the ambiguous polarity feature resembles less a clear streak and more a collection of streaks close enough together that it is difficult to distinguish between them.
As we almost always observe such features to be multipolar, we do not count these as faculae.

Any uncertainty introduced by this process and the assumptions we make here would likely be towards over-counting the number of PFe, as we would count some streaks as faculae that are in fact multipolar regions.
Method 2 overlays the magnetogram on the progressive standard deviation for each time sample and can therefore be used to distinguish between unipolar and multipolar regions throughout the day under consideration. 
However, it becomes difficult to do this on a large enough scale to provide statistical information about the behavior of faculae, as we do in this work, because the movie rapidly becomes visually confusing as the day progresses.

We consider automating the counting process using these methods to be the most promising extension of this work.
An automated counting method holds multiple advantages over a manual one.
It would allow a significant scaling-up of the method, providing counts of all days of the SDO mission, because it would be limited only by the amount of data rather than by the time it takes for a human to manually count the streaks.
It would also increase the objectivity in the facular-identification process, as the algorithm would follow a consistent set of rules to a degree not possible to humans.
We have tried to minimize subjectivity by counting in a randomized, blind ordering, but an automated counting process should do even better.
Further, an automated method would be able to work with a quantity of data that humans cannot.
When counting the streaks using the first method, we only overlay the magnetogram from 12:00Z.
The magnetic-field data over the whole day would provide far more information but is difficult for a human to visualize, because overlaying more than one frame of the magnetogram would start to obscure the data, making it difficult or impossible to count the streaks.
We developed the second method, the progressive standard-deviation movie, for this purpose, but while it is a better way to associate faculae with the magnetic field, it is more difficult to measure the positions of the faculae.
An automated counting algorithm, however, could consider both data series, HMI${}_\mathrm{IC}$ and HMI${}_\mathrm{M}$, over time without any loss of quality or clarity in the PFe measurements.

\section{Conclusions}\label{sec:conclusions}

We present two new methods for measuring the properties of polar faculae: a daily standard-deviation image and a movie of the progressive standard deviation overlain with the magnetic field.
These methods produce facular counts whose solar-cycle dependence agrees with previous studies and that are correlated with the strength of the polar magnetic field.

Further, these methods provide an improvement over previous methods.
They make faculae more visible and provide assurance that a bright point is a facula rather than a statistical brightening by tracking it for many images.
They allow for observation of all faculae over the course of a day, rather than only an instantaneous count.
They provide a method and visualization for tracking the evolution of PFe over the course of a day.
They enable distinguishing between unipolar faculae and multipolar regions.

They allow, for the first time, reliable measurement and a statistical distribution of facular lifetime for a large number of PFe.
From our analysis of 9291 faculae, we find a median lifetime of 4.9 hours.
One possible extension of this research would be to calculate the lifetime as a function of the phase of the solar cycle.

While both methods increase our understanding of PFe, each has purposes to which it is better suited.
The first method, the standard-deviation image, provides all facular data for a day in a single image, making it easy to count PFe and measure their positions and lifetimes.
The second method, the progressive standard-deviation movie, shows how faculae and their associated magnetic-field elements evolve and develop over time, providing a useful visualization for understanding the evolution of these features.

While we use these methods to track PFe only while they are visible, observations of the area a facula occupies for a full disk passage could confirm whether the underlying magnetic structure that forms a facula is present for longer than a single brightening, as concluded by \cite{Hirayama:1978PASJ...30..337H}.

It is possible to automate the identification of polar faculae in the daily images of standard deviation and the measurement of their lifetime. This would allow daily measurements over almost the entirety of Solar Cycle 24 and at least part of Solar Cycle 25 and a quantitative comparison with the polar-field values from HMI and WSO as well as comparisons with the PF counts from the early stages of Solar Cycle 24 from \citet{munoz-jaramillo_calibrating_2012} and all of Solar Cycle 24 from \cite{janssens_prediction_2021}.

These methods would not be possible without the high resolution, frequent imaging, and stable pointing of SDO's HMI instrument, as well as the lack of seeing experienced by SDO in its inclined geosynchronous orbit.

\begin{acks}
This research was supported, in part, by the SDO Project. The HMI data is courtesy of the SDO HMI Science Investigation Team.
Some of the computing was performed at the JSOC, which is located at the Stanford University and operated by the HMI Science Investigation Team.
This research made use of version 4.2 of Astropy (http://www.astro-\\py.org), a community-developed core Python package for Astronomy \citep{astropy:2013, astropy:2018}.
This research used version 2.1.3 of the SunPy open source software package \citep{sunpy2_1_3,sunpy_community2020}.
This work was performed while B. H.-A. was a research support assistant at the Catholic University of America.
\end{acks}

\bibliographystyle{spr-mp-sola}
\bibliography{bibliography} 

\end{article} 
\end{document}